\documentclass[conference]{IEEEtran}
\usepackage{multirow}
\usepackage{subcaption}
\usepackage{float}
\usepackage{amsmath}
\usepackage{graphicx}
\usepackage{balance}
\usepackage{hyperref}
\usepackage{xcolor}
\usepackage{comment}
\usepackage[skip=-2pt]{caption}
\begin{document}

\title{A Label Correction Algorithm Using Prior Information for Automatic and Accurate Geospatial Object Recognition}

\makeatletter
\newcommand{\linebreakand}{%
  \end{@IEEEauthorhalign}
  \hfill\mbox{}\par
  \mbox{}\hfill\begin{@IEEEauthorhalign}
}
\makeatother

\author{Weiwei Duan\\
University of Southern California\\
{\tt\small weiweiduan@usc.edu}

\and
Yao-Yi Chiang\\
University of Minnesota\\
{\tt\small yaoyi@umn.edu}

\and
Stefan Leyk\\
University of Colorado Boulder\\
{\tt\small stefan.leyk@colorado.edu }
 \linebreakand
\and
Johannes H. Uhl\\
University of Colorado Boulder\\
{\tt\small johannes.uhl@colorado.edu}

\and
Craig A. Knoblock\\
University of Southern California\\
{\tt\small knoblock@isi.edu}
}
\maketitle
\begin{abstract}
% why this is an interesting problem
Thousands of scanned historical topographic maps contain valuable information covering long periods of time, such as how the hydrography of a region has changed over time. 
% problem statement
Efficiently unlocking the information in these maps requires training a geospatial objects recognition system, which needs a large amount of annotated data. Overlapping geo-referenced external vector data with topographic maps according to their coordinates can annotate the desired objects' locations in the maps automatically. However, directly overlapping the two datasets causes misaligned and false annotations because the publication years and coordinate projection systems of topographic maps are different from the external vector data. 
% my solution
 We propose a label correction algorithm, which leverages the color information of maps and the prior shape information of the external vector data to reduce misaligned and false annotations. 
% what follows the solution
The experiments show that the precision of annotations from the proposed algorithm is 10\% higher than the annotations from a state-of-the-art algorithm. Consequently, recognition results using the proposed algorithm's annotations achieve 9\% higher correctness than using the annotations from the state-of-the-art algorithm. 

\end{abstract}

\vspace{-5pt}
\section{Introduction}
\vspace{-2pt}
Detecting objects' pixels in scanned historical map images can extract geospatial objects' locations, which help study the evolution of natural features and human activities~\cite{intro:maps1,intro:maps2,intro:maps3,intro:maps4}. The paper proposes a fully automatic system to extract geospatial objects' locations in the maps. Our system removes the manual annotation work for training a pixel-wise classifier by leveraging the external geo-referenced (often vector) data. For example, the green lines in Figure~\ref{fig:false_annotations} are external vector data after aligning them with the map. The desired objects' annotations are the intersection areas between the vector (green) lines and the maps. However, the annotations from the vector data have misaligned and false annotations. The misaligned annotations (Figure~\ref{fig:false_annotation_railroads}) typically result from different scales and coordinate projection systems between the external data and the maps. The false annotations in Figure~\ref{fig:false_annotation_waterlines} result from different publication year between the vector data and maps. The vector data represents canals built after the map publishing year. The misaligned and false annotations could mislead a semantic segmentation model to extract inaccurate object locations in the maps.

Existing methods deal with misaligned and false annotations from the external datasets before or during the training process. During the training process, the existing methods design new loss functions~\cite{entropy_loss,feature_sim_loss,normal_cut_loss_road}, which exploit the information from the image itself instead of only relying on annotations to classify pixels. For example, the roads detection model~\cite{normal_cut_loss_road} using OpenStreetMap (OSM) for annotations includes a new loss function: normalized cut loss (Ncut), which calculates the similarities among the annotated pixels in the color (RGB) and spatial (XY) spaces. The true roads' pixels should have similar colors and be close to each other. To maximize the similarity among the annotated road pixels, Ncut classifies falsely annotated pixels into the non-road category. However, the similarity in the color and spatial space is not reliable when the false annotations are not minor and similar to each other. Correcting the annotations before the training process handles the non-minor annotation noise. For example, a recent vector-to-raster algorithm~\cite{alignment} moves the vector data to an area, which has color in the target objects' color range in the maps. However, some pixels surrounding the target objects in the maps may also be in the color range. The vector-to-raster algorithm often aligns the vector data to the surrounding area of the target objects. Consequently, aligned vector data falsely annotates pixels surrounding the target objects in the maps.

\begin{figure}[h]
\captionsetup[subfigure]{labelformat=empty}
\centering
    \subfloat[\small]{
    \label{fig:false_annotation_railroads}
    \frame{\includegraphics[angle=90,width=0.4\linewidth]{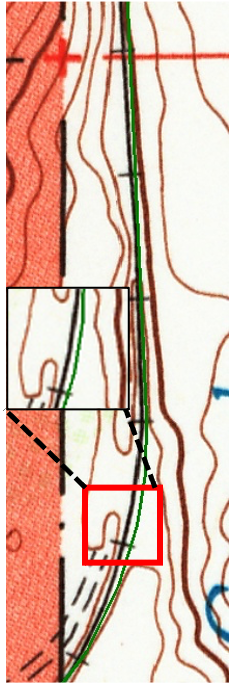}}}
    \hspace{0.005cm}
    \subfloat[\small]{
    \label{fig:false_annotation_waterlines}
    \frame{\includegraphics[width=0.45\linewidth]{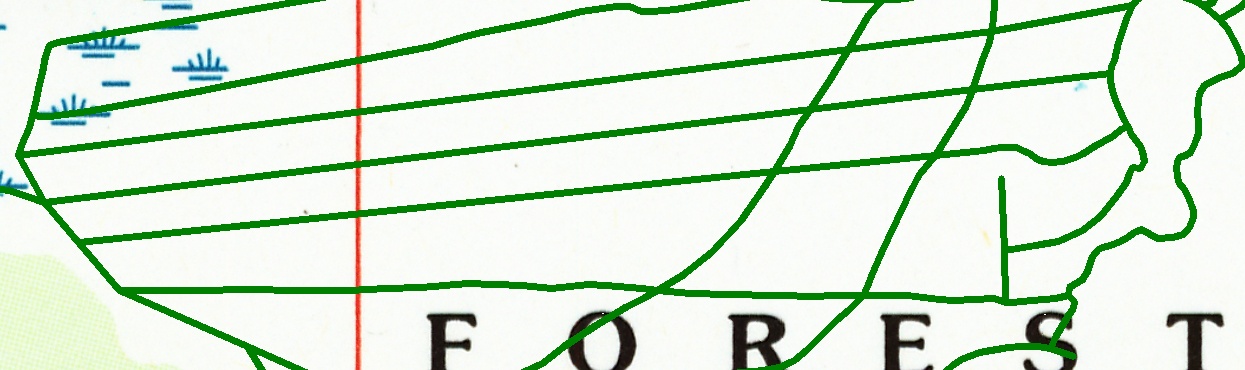}}}
    \caption{\small The green lines are vector data after aligning with the map. The left figure shows the misaligned annotations for railroads. We magnify the misalignment in the red box. The right figure shows the false annotations for waterlines. }
    \label{fig:false_annotations}
\end{figure}

We propose a label correction algorithm (LCA) to reduce misaligned and false annotations before the training process in our linear geospatial object recognition system. 
\begin{comment}
Unlike the new loss solutions, our system can handle different levels of annotation noises.
\end{comment}
Figure~\ref{fig:workflow} shows the workflow of our system. LCA in the training process (the yellow box) aims to find a group of foreground pixels. The foreground pixels have similar colors. The shape of the foreground pixels group is similar to the target object's shape. The target object's shape is the prior information provided by the vector data. We design the objective function in LCA to find a color homogeneous pixel group, which has a similar shape to the target object's shape. The group of foreground pixels from LCA are the annotations for the target objects. LCA corrects the misaligned annotations by finding the foreground pixel group. LCA removes the false annotations when LCA does not find foreground pixels. Figure~\ref{fig:false_annotation_waterlines} shows that the false annotations annotate background pixels as target objects. After training the semantic segmentation model, the inference process (the blue box) detects the target objects' pixels in the maps. Finally, the system converts the recognition results into vector data.

\begin{figure}[h]
	\includegraphics[width=1.0\linewidth]{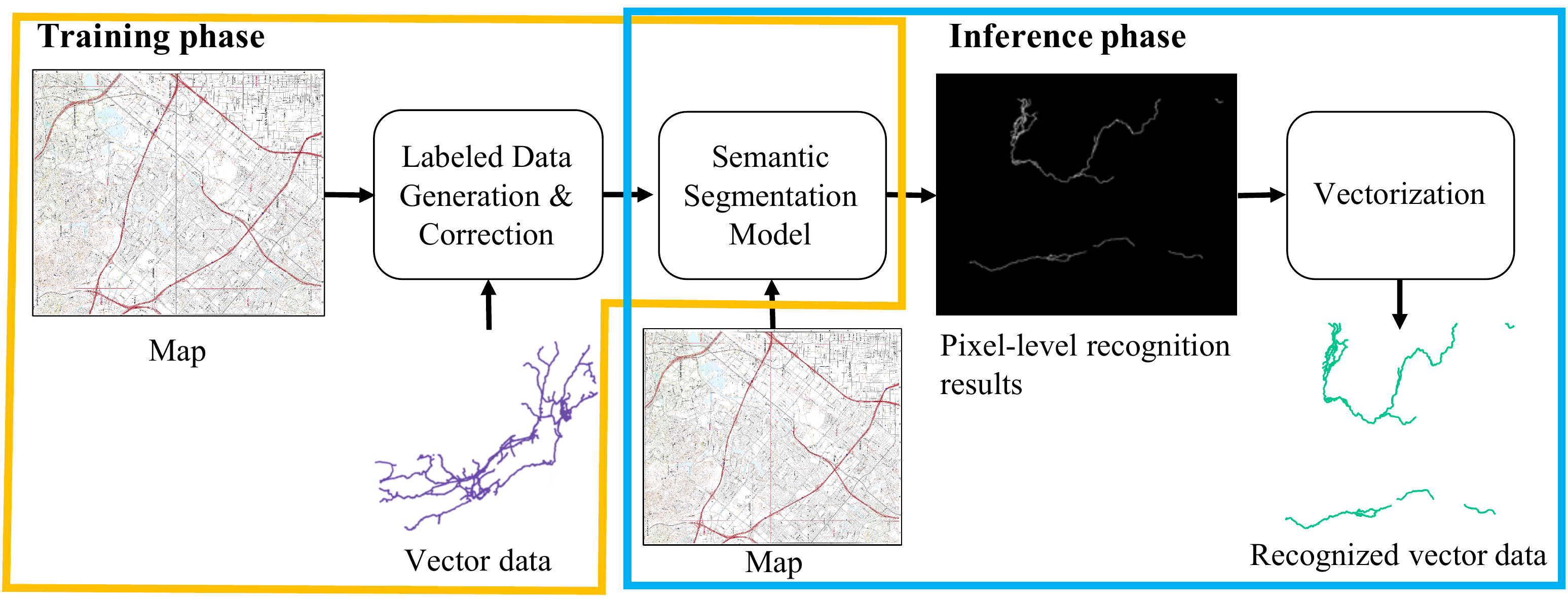}
	\caption{The workflow of proposed object recognition system.}
	\label{fig:workflow}
\end{figure}

The proposed linear geospatial object recognition system automatically extracts the linear geospatial objects' locations in the scanned topographic maps. The proposed LCA reduces the misaligned and false annotations by using the color information of maps and prior shape information of the external vector data. The experiment section shows that the annotations from the LCA are more accurate than the baselines. Consequently, the semantic segmentation models trained using the annotations from the LCA obtained better recognition results than the state-of-the-art methods.

\vspace{-0.3cm}
\section{Geospatial Object Recognition System}
\vspace{-0.1cm}
This section explains the three major components in the proposed geospatial object recognition system: labeled data generation and correction, semantic segmentation model for detecting target objects' pixels in the maps, and vectorization for converting the recognition results into the vector data. 

\vspace{-0.3cm}
\subsection{Labeled Data Generation}
\label{sec:label_gen}
\vspace{-0.1cm}
Labeled data generation in our system has three steps. First, our system creates the buffer zone around the vector data. The buffer size is the target objects' width in the maps. Second, our system rasterizes the buffered vector data into a white-and-black mask image. The white areas represent the target objects' locations in the maps. Third, our system annotates white pixels as the target object pixels in the maps. 

\vspace{-0.3cm}
\subsection{Label Correction}
\vspace{-0.1cm}
The proposed LCA is a variant of the Chan-Vese algorithm. We first briefly introduce the Chan-Vese algorithm. 

\subsubsection{Chan-Vese Algorithm}
The Chan-Vese (CV) algorithm~\cite{cv} segments the pixels in an image into the foreground and background areas. The CV's input is an image. The output is a binary image, in which white and black areas are foreground and background, respectively. The CV's objective function calculates the color homogeneity within the foreground and background area. The CV updates the pixels' assignment until the objective function converges to a minimum. The objective function is as follows:
\vspace{-0.1cm}
\begin{align}\label{eq:cv_eq}
&E(c_{1}, c_{2}, \phi)=\int_{\Omega}\{(u(x,y)-c_{1})^2H(\phi(x,y)) \nonumber \\
&+(u(x,y)-c_{2})^2(1-H(\phi(x,y)))\}dxdy 
\end{align}

\noindent where $u(x,y)$ is the pixel color at coordinate (x,y) in an image ( $\Omega \rightarrow R^2$).  $c_{1}$ and $c_{2}$ are two scalar variables representing the average color of foreground and background areas, respectively. $\phi(x,y)$, a Lipschitz function~\cite{cv}, represents the foreground-and-background segmentation. $H(\phi(x,y))$, the Heaviside function~\cite{cv}, represents the binary image, the CV's output. Ones in $H(\phi(x,y))$ represent foregrounds. 

In each iteration, the updates of $c{1}$, $c_{2}$, and $\phi(x,y)$ are:
\begin{align}\label{eq:c1_update}
    c_{1}=\frac{\int_{\Omega}u(x,y)H(\phi(x,y))dxdy}{\int_{\Omega}H(\phi(x,y))dxdy}
\end{align}
\vspace{-0.3cm}
\begin{align}\label{eq:c2_update}
    c_{2}=\frac{\int_{\Omega}u(x,y)(1-H(\phi(x,y)))dxdy}{\int_{\Omega}(1-H(\phi(x,y)))dxdy}
\end{align}
\vspace{-0.3cm}
\begin{align}\label{eq:phi_update}
    \frac{\partial\phi(x,y)}{\partial t}=-[(u(x,y)-c_{1})^2-(u(x,y)-c_{2})^2]\delta(\phi(x,y))
\end{align}

\noindent where $\delta(\phi(x,y))$ is the derivation of the Heaviside function $H(\phi(x,y))$. In practice, the CV ignores $\delta(\phi(x,y))$ to accelerate the optimization process. Song et al.~\cite{cv_fast_backprop} provide detailed proof about the correctness of fast optimization of the CV. In each iteration, the CV algorithm first updates $c_{1}$ and $c_{2}$ according to eq.~\ref{eq:c1_update} and~\ref{eq:c2_update}. Secondly, the CV algorithm sets $\phi(x,y)$ to one if $(u(x,y)-c_{1})^2\leq(u(x,y)-c_{2})^2$. The CV minimizes the objective function eq.~\ref{eq:cv_eq} by updating $c{1}$, $c_{2}$, and $\phi$ until the objective function does not change.

\subsubsection{Label Correction Algorithm (LCA)}
% The overview of LCA, what the inputs, outputs, the goal
LCA aims to find the desired group of foreground pixels in the images. The desired group of foreground pixels has a similar shape to the target object's shape. The desired foreground pixels' locations are the target object's annotations in the maps. The the target object's shape is the prior information, which is from the external vector data after an affine transformation. Because of the different scales and coordinate projections between the vector data and the maps, the vector data's shape and the target object are not exactly the same. LCA uses the affine transformation to approximately transform the shape of the vector data to the target objects in the maps. 

The inputs for LCA are image, prior shape, and pixels-of-interest (PoI). PoI, a binary image, provides candidates for desired foreground pixels. Ones in PoI are candidate pixels. A map might have more than one group of pixels, which have similar shape as the target object's shape. PoI reduces the search space for the desired foreground pixels. Since the misaligned vector data often partially overlaps with or is several pixels away from the target objects in the maps, LCA uses a large buffer to ensure that the PoI contains the desired foreground pixels. Another input, the prior shape, is also a binary image derived from the external vector data. Section~\ref{sec:label_gen} describes how to generate the prior shape. Ones in the binary image represent the shape prior. The shape of desired foreground pixels is similar to the target object's shape. The shape prior provides the target object's shape after the affine transformation.

LCA has two optimization goals to find the desired foreground pixels: 1. grouping the images into the homogeneous color foreground and background areas. 2. maximizing the intersection between the desired foreground pixels and the target object's shape. The intersection area size represents the similarity between the desired foreground pixels and the target object's shape. The objective function in LCA representing the two goals is:
\begin{align}\label{eq:cvshp_eq}
    &E(c_{1},c_{2},\phi,L,\psi)= \\ \nonumber
    &\int_{\Omega}(u-c_{1})^2H(\phi)+(u-c_{2})^2(1-H(\phi))dxdy\\ \nonumber
    &+\lambda\int_{\Omega}(H(\phi)H(L)-H(\psi))^2dxdy
\end{align}

\noindent where $u$, $\phi$, $L$, $\psi$ are Lipschitz functions~\cite{cv} for the image, the foreground-and-background segmentation, the PoI, and the target object's shape, respectively. $H(\cdot)$, a Heaviside function, represents a binary image. Ones in $H(\phi)H(L)$ are the desired foreground pixels. $c_{1}$, $c_{2}$ are scalar variables representing the average color of foreground and background areas, respectively. The first term on the right-hand side of eq.~\ref{eq:cvshp_eq} is for the homogeneous color goal. The second term in the right hand side of eq.~\ref{eq:cvshp_eq} calculates the intersection between the desired foreground pixel and the target object's shape. Large intersection area indicates the high shape similarity between the desired foreground pixels and the target object. 

The relationship between the prior shape and target object's shape via the affine transformation parameters is:
\begin{align}\label{eq:affine}
    \psi(x,y)=s\psi_{0} \Bigg[ \Bigg.  \mbox{affine}(x^{*}, y^{*}), \mbox{affine}(x^{*}, y^{*}) 
    \Bigg. \Bigg]
\end{align}
\vspace{-0.5cm}
\begin{align}\label{eq:affine_x}
    \mbox{affine}(x^{*}, y^{*})= \frac{1}{r} \Bigg[ \Bigg. 
    &(x^{*}-tr_x)(\cos{\theta}+sh_{y}\sin{\theta})\nonumber\\
    &+(y^{*}-tr_y)(\sin{\theta}+sh_x\cos{\theta})\Bigg. \Bigg]
\end{align}
\vspace{-0.5cm}
\begin{align}\label{eq:affine_y}
   \mbox{affine}(x^{*}, y^{*})= \frac{1}{r} \Bigg[ \Bigg. &(x^{*}-tr_x)(-\sin{\theta}+sh_y\cos{\theta})\nonumber\\
   &+(y^{*}-tr_y)(\cos{\theta}-sh_x\sin{\theta})
    \Bigg. \Bigg]
\end{align}

\noindent where $\psi_{0}$ and $\psi$ are the prior shape and target object's shape, respectively. The affine transformation parameters are the translation along the x-axis ($tr_x$), the translation along the y-axis ($tr_y$), the scale ($s$), the rotation angle ($\theta$), the shear along the x-axis ($sh_x$), and the shear along the y-axis ($sh_y$). 

The optimization process minimizes the objective function in eq.~\ref{eq:cvshp_eq} by updating pixels into foreground and background areas ($\phi$), the PoI ($L$) and the target object's shape ($\psi$). The detailed updates are:
\vspace{-0.2cm}
\begin{align}\label{eq:phi_shp_update}
    \frac{\partial\phi}{\partial t}=&-[(u-c_{1})^2-(u-c_{2})^2 \nonumber\\
    &+2\lambda H(L)(H(\phi)H(L)-H(\psi))]\delta(\phi)
\end{align}
\vspace{-0.5cm}
\begin{align}\label{eq:l_update}
    \frac{\partial L}{\partial t}=-2\lambda H(\phi)[H(\phi)H(L)-H(\psi)]\delta(L)
\end{align}
\vspace{-0.5cm}
\begin{align}\label{eq:psi_update}
    \frac{\partial\psi}{\partial t}=-2\lambda[H(\phi)H(L)-H(\psi)]\delta(\psi)
\end{align}

In each optimization iteration, LCA first updates $c_{1}$ and $c_{2}$ according to eq.~\ref{eq:c1_update} and~\ref{eq:c2_update}. Secondly, LCA updates $\phi$, $L$ and $\psi$ according to the equations above. In practice, LCA also ignores $\delta(\phi)$, $\delta(\psi)$, and $\delta(L)$ to accelerate the optimization process like the CV algorithm. LCA sets $\phi(x,y)$ to one (i.e., setting the pixel as the foreground), if $-[(u-c_{1})^2-(u-c_{2})^2+2\lambda H(L)(H(\phi)H(L)-H(\psi))] \geq 0$. The weight $\lambda$ makes $[H(L)(H(\phi)H(L)-H(\psi)]$ and $[(u-c_{1})^2-(u-c_{2})^2]$ at the same magnitude. For the update of PoI ($L$) and the target object's shape ($\psi$), we follow the optimization strategy proposed by Peng et al.~\cite{local-level-set}. LCA only updates $L(x,y)$ and $\psi(x,y)$ where the initial $L(x,y)>0$ and $\psi(x,y)>0$, respectively. In other words, the size of the PoI and target object's shape can be smaller than or the same as the original size during the optimization process. LCA sets $L(x,y)$ to one if $H(\phi)[H(\phi)H(L)-H(\psi)] \geq 0$. For updating $\psi(x,y)$, LCA first updates the affine transformation parameters. Secondly, LCA uses the updated affine transformation parameters to move the shape prior towards desired foreground pixels. The transformed shape prior is the target object's shape ($\psi$). Thirdly, LCA sets $\psi$ to one if $[H(\phi)H(L)-H(\psi)] \geq 0$.

The following equations describe the updates affine transformation parameters in detail:
\begin{align}\label{eq:dx_update}
    \frac{\partial tr_x}{\partial t} = &\int_{\Omega}-(H(\phi)H(L)-H(\psi)) \nonumber\\
    &\{\psi_{0x}(x^*, y^*)(\cos{\theta}+sh_{y}\sin{\theta}) \nonumber\\
    &-\psi_{0y}(x^*, y^*)(\sin{\theta}+sh_{x}\cos{\theta}\}\delta(\phi)dxdy
\end{align}
\vspace{-0.6cm}
\begin{align}\label{eq:dy_update}
    \frac{\partial tr_y}{\partial t} = &\int_{\Omega}-(H(\phi)H(L)-H(\psi)) \nonumber\\ 
    &\{\psi_{0x}(x^*, y^*)(\sin{\theta}+sh_{x}\cos{\theta}) \nonumber\\
    &+\psi_{0y}(x^*, y^*)(\cos{\theta}-sh_x\sin{\theta})\}\delta(\phi)dxdy
\end{align}
\vspace{-0.6cm}
\begin{align}\label{eq:scale_update}
    \frac{\partial s}{\partial t} = &\int_{\Omega}-(H(\phi)H(L)-H(\psi))\{-\psi_{0}(x^*, y^*) \nonumber\\
    &+\psi_{0x}(x^*, y^*)x^* + \psi_{0y}(x^*, y^*)y^*\}\delta(\phi)dxdy
\end{align}
\vspace{-0.6cm}
\begin{align}\label{eq:rotation_update}
    \frac{\partial \theta}{\partial t} = &\int_{\Omega}-(H(\phi)H(L)-H(\psi))\{-s\psi_{0x}(x^*, y^*)y^*) \nonumber\\
    &+s\psi_{0y}(x^*, y^*)x^*\}\delta(\phi)dxdy
\end{align}
\vspace{-0.6cm}
\begin{align}\label{eq:shearx_update}
    \frac{\partial sh_x}{\partial t}=
    \int_{\Omega}&-(H(\phi)H(L)-H(\psi))\{\psi_{0x}(x^*, y^*)\nonumber\\
    &((x-a)\cos{\theta}+(y-b)\sin{\theta})\}\delta(\phi)dxdy
\end{align}
\vspace{-0.6cm}
\begin{align}\label{eq:sheary_update}
    \frac{\partial sh_y}{\partial t}=\int_{\Omega}&-(H(\phi)H(L)-H(\psi))\{\psi_{0x}(x^*, y^*) \nonumber\\ 
    &((x-a)\cos{\theta}+(y-b)\sin{\theta})\}\delta(\phi)dxdy
\end{align}

\noindent where $\psi_{0x}$, and $\psi_{0y}$ are partial derivation of $\psi_{0}$ of x and y, respectively, described as follows:
\vspace{-0.2cm}
\begin{align}
    \psi_{0x} = \frac{\partial \psi_{0}}{\partial x}, \psi_{0y} = \frac{\partial \psi_{0}}{\partial y}
\end{align}

\subsection{Semantic Segmentation Model}
Our system uses deeplabv3+~\cite{deeplabv3+} as the semantic segmentation model. The deeplabv3+ is the champion of the PASCAL VOC 2012 test~\cite{pascal-voc-2012} for semantic segmentation in 2018. During the training phase, deeplabv3+ has two inputs: 1. the scanned topographic maps, 2. the annotations for the target objects, which are the outputs from LCA. The learning task for deeplabv3+ is detecting the target objects' pixels in the maps. During the inference phase, deeplabv3+ classifies the pixels into the target category and non-target category. 
\begin{comment}
In the introduction, we mentioned several new losses dealing with noisy annotations, such as the normalized cut loss. We can add the new loss as the extra loss for deeplabv3+.
\end{comment}

\vspace{-0.1cm}
\subsection{Vectorization}
% \vspace{-0.1cm}
The last step in our system is converting the pixel-level recognition results into vector data. Our system exploits the vectorization method proposed in the SpaceNet Road Detection and Routing Challenge.~\footnote{\url{https://github.com/SpaceNetChallenge/RoadDetector/blob/master/albu-solution/src/skeleton.py}} The vectorization method first skeletonizes the pixel-level recognition results. Secondly, the method finds the end and turning nodes in the skeleton results. Third, the method uses edges to link the nodes.
\vspace{-3pt}
\section{Experiment and Analysis}
 
\subsection{Experiment Data and Settings}
% In this section, we first introduce the experiment data. Secondly, we explain the baselines and our method in the labeled data generation and correction step. Thirdly, we introduce the training details for the semantic segmentation model. At last, we explain the evaluation metrics. 

\paragraph{\textbf{Experiment Data}} We tested two target linear objects: railroads and waterlines, in the scanned topographic maps, covering Bray in California published in 2001, and Louisville in Colorado published in 1965, from the United States Geological Survey (USGS). The maps contain diverse geospatial objects, including railroads, waterlines, roads, lakes, mountains, and wetlands. The dimensions of the two map sheets are large: the Bray map is 12,943-pixel height and 16,188-pixel width, and the Louisville map is 11,347-pixel height 13,696-pixel width. The external vector data for annotation is the vector-formatted map data from USGS published in 2018. 

\paragraph{\textbf{Labeled Data Generation \& Correction}} We generated three groups of labeled data. The first two groups are the baselines, and the other is from the proposed algorithm, LCA.
\begin{itemize}
    \item \textbf{Group one (original vector)}: Our system generates the annotations from USGS vector data directly following the steps described in Section~\ref{sec:label_gen}. 
    \item \textbf{Group two (vector-to-raster)}: Our system first uses the vector-to-raster algorithm~\cite{alignment} to align original USGS vector data to the target objects in the maps. Secondly, our system generates the annotations from the aligned vector data following the steps described in Section~\ref{sec:label_gen}. The buffer size for the first two groups is 5-pixel, 5-pixel, and 3-pixel for railroads, waterlines in the Bray map, and railroads in the Louisville map, respectively. 
    \item \textbf{Group three (LCA)}: The annotations are from LCA. The buffer for PoI in LCA is 30-pixel, which is sufficient to cover the areas including target objects in the maps. We set $\lambda$ in eq.~\ref{eq:cvshp_eq} as 3.
    \end{itemize}

\paragraph{\textbf{Semantic Segmentation models}} Our system uses deeplabv3+ as the semantic segmentation model. We tested on two groups of loss functions for deeplabv3+, one with the normalized cut loss (named as \textbf{Ncut}) and the other without the normalized cut loss (named as \textbf{no-Ncut}). The Ncut loss~\cite{normal_cut_loss_road, normal_cut_loss_seg} is an existing method dealing with the label noises. We trained the two semantic segmentation models using three sets of annotations for each target object. 

\paragraph{\textbf{Training Data for the Models}}Due to the large dimensions of the test maps, our system cropped maps to generate training samples as 128-*128-pixel. Our system cropped 1,000 training samples containing the target objects and 3,000 training samples not containing the target objects. Due to the diverse contents in the maps, our system generated more non-target samples than target samples. 

\paragraph{\textbf{Training Details for the Models}} We used Adam as the optimizer. For the model without the normalized cut loss,  the learning rate was 2e-4. For the model with the normalized cut loss, the weight for the normalized cut loss is 1e-3, and the learning rate was 1e-4.

\paragraph{\textbf{Evaluation Metrics: }} For the training data generation step, we used the pixel-level precision, recall, and $F_{1}$ score to evaluate the quality of annotations. We evaluated the recognized vector lines using correctness, completeness~\cite{correct_complete}, and the Average Path Length Similarity (APLS), which are commonly used metrics for linear geospatial objects. Correctness represents how much length of the recognized lines is the true target objects. Completeness represents how much length of the true target objects is recognized. APLS~\cite{APLS}, scaling from zero to one, describes the linear objects' continuity. Low APLS indicates that the recognition results for continuous lines have many gaps. We manually digitized railroads and waterlines in the maps to generate vector data as the ground truth.

\subsection{Annotations Results and Analysis}

\begin{table*}[h]
\resizebox{2\columnwidth}{!}{\begin{tabular}{|c|c|c|c|c|c|c|c|c|c|}
\hline
\multirow{2}{*}{} & \multicolumn{3}{c|}{Bray railroads} & \multicolumn{3}{c|}{Louisville railroads} & \multicolumn{3}{c|}{Bray waterlines} \\ \cline{2-10} 
& Precision   & Recall    & $F_{1}$        & Precision     & Recall      & $F_{1}$         & Precision    & Recall    & $F_{1}$        \\ \hline
Original vector                   & 37.02\%     & 58.72\%   & 45.41\%   & 59.20\%       & 60.31\%     & 59.75\%     & 65.01\%      & 96.44\%   & 77.67\%   \\ \hline
Vector-to-raster                    & 77.78\%     & \textbf{87.85\%}   & 82.55\%   & 96.56\%       & \textbf{83.73\%}     & 89.69\%     & 86.24\%      & 88.65\%   & 87.43\%   \\ \hline
LCA                  & \textbf{95.17\%}     & 86.15\%   & \textbf{90.42\%}   & \textbf{99.78\%}       & 83.02\%     & \textbf{90.63\%}     & \textbf{96.39\%}      & \textbf{88.98\%}   & \textbf{92.53\%}   \\ \hline
\end{tabular}}
\caption{\label{tab:alignment} The evaluation for the quality of training data annotations using pixel-level precision, recall and $F_{1}$ score.}
\end{table*}

The precision in Table~\ref{tab:alignment} shows that LCA improved the annotation quality compared with the annotations from the other two groups. The average precision of annotations is around $10\%$ higher than the rest two groups of annotations. Figure~\ref{fig:alignemnt_example} from top to bottom shows railroads' annotations in the Bray map from the three groups. The pixels within the blue outlines are the annotated railroad pixels. Annotations from the original vector and vector-to-raster groups show that the misalignment between the vector data and the railroads' center lines in the maps causes the background area surrounding the railroads to be falsely annotated as railroads. In contrast, LCA segmenting color homogeneous foreground and background areas reduces the label noises in the surrounding areas. 

The average recall from LCA is $85\%$. The external vector data averagely missed 14.84\% of the target objects in the maps because the target objects disappeared when the vector data was published. The missing annotations from the vector data are the primary reason for the $85\%$ recall. Another reason is that LCA falsely annotates other objects in the maps when the vector data is closer to other objects than the target objects. For example, the area within the red outline in Figure~\ref{fig:bad_annotation_exmple_align_ijgis} shows that LCA falsely annotated roads as railroads in the Bray map. The yellow line (vector line) is closer to the road than the railroad. However, 85\% recall indicates that the misaligned original vector data is sufficiently close to the target objects in most locations in the maps.

\begin{figure}[h]
    \centering
    \subfloat[The areas within the blue outlines are the annotated railroads (target) objects in the Bray map.]{
    \label{fig:alignemnt_example}
	\includegraphics[width=0.4\textwidth]{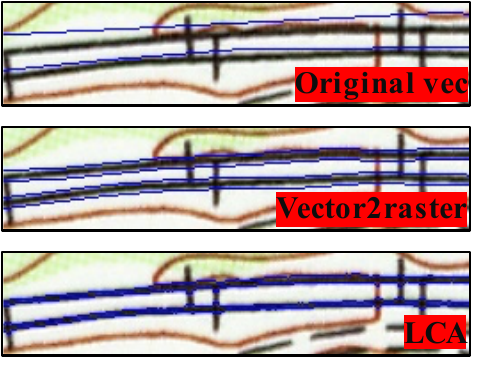}}

	\centering
	\subfloat[The red outline shows the false annotations from LCA. The yellow line is the original vector data. LCA falsely annotates railroad locations when the vector data is closer to other objects than the railroads.]{
	\label{fig:bad_annotation_exmple_align_ijgis}
	 \frame{\includegraphics[width=0.38\textwidth]{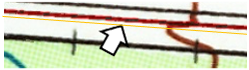}}}
	 \caption{Annotations visualization}
\end{figure}

\begin{table*}[h]
\resizebox{\textwidth}{!}{\begin{tabular}{|c|c|c|c|c|c|c|c|c|c|c|}
\hline
\multirow{2}{*}{}                    &         & \multicolumn{3}{c|}{Bray railroads} & \multicolumn{3}{c|}{Bray waterlines} & \multicolumn{3}{c|}{Louisville railroads} \\ \cline{3-11}
 &         & Original & Vec2raster  &LCA & Original  & Vec2raster     & LCA       & Original         & Vec2raster          & LCA        \\ \hline
\multirow{2}{*}{Correctness}  & Ncut  & 0.4437 & 0.2385 & 0.7915 & 0.0 & 0.7429     & 0.6743     & 0.1139       & 0.3230       & 0.5034      \\ \cline{2-11} 
& no-Ncut & 0.4937     & 0.7644     & \textbf{0.8078} & 0.7775 & 0.7968 & 0.8223 & 0.6084 & 0.7259  & \textbf{0.8661}      \\ \hline
\multirow{2}{*}{Completeness} & Ncut  & 0.4772 & 0.5025 & 0.6482    & 0.0        & 0.8524     & 0.9341     & 0.2167       & 0.4945 & 0.2704 \\ \cline{2-11} 
& no-Ncut & 0.4543 & 0.7918 & \textbf{0.8408} & 0.9454     & 0.8874     & \textbf{0.9671}     & 0.4664       & \textbf{0.8182}       & 0.7154      \\ \hline
\multirow{2}{*}{APLS}         & Ncut  & 0.3319            &  0.2437          & \textbf{0.6353} & 0.0 &0.2880  & \textbf{0.5424}  & 0.1239     & 0.1799             & 0.1704            \\ \cline{2-11} 
& no-Ncut & 0.3120     & 0.4052     & 0.5750    & 0.3533            & 0.3142           & 0.4600     & 0.5333             & 0.3059             & \textbf{0.4993} \\ \hline
\end{tabular}}
\caption{\label{tab:recognition} The evaluation for the vectorized recognition results. "Original", "Vec2raster", and "LCA" represent three sets of annotations. "Ncut" and "no-Ncut" represent deeplab v3+ with and without the normalized cut loss.}
\end{table*}

\subsection{Recognition Results and Analysis}

In this section, first, we use deeplabv3+ without the normalized cut loss to analyze the impact of annotation quality on the recognition results. Second, we compare the recognition results from deeplabv3+ with and without the normalized cut loss to verify if the normalized cut loss is suitable for our label noise problem.

\subsubsection{The Impact of Annotation Quality} 
The correctness from deeplabv3+ without the normalized cut loss in Table~\ref{tab:recognition} shows that the accurate annotations from LCA obtain the highest correctness compared to the results from the other two groups of annotations. Figure~\ref{fig:dlab_bray_railroads_detailed_results}, ~\ref{fig:dlab_bray_waterlines_extraction} and ~\ref{fig:dlab_louisville_railroads_extraction} show the extraction comparisons for railroads, waterlines in the Bray map and railroads in the Louisville map, respectively. The rows from top to bottom represent the results from the original vector, vector-to-raster, and LCA group, respectively. Green and red lines represent the true positive and false positive recognition, respectively. The results from LCA have more true positive lines than the original vector group and have fewer false positive lines than the vector-to-raster group. The completeness in Table ~\ref{tab:recognition} shows that deeplabv3+ trained using the LCA's annotations extracted $5\%$ and $8\%$ longer railroads and waterlines in the Bray maps than the vector-to-raster group, respectively. The recognized (green) lines in Figure ~\ref{fig:dlab_tp_bray_waterlines_orig_ijgis} and ~\ref{fig:dlab_tp_bray_waterlines_align_ijgis} show that the recognized waterlines from LCA are longer than the vector-to-raster group.

\begin{figure}[h]
    % \captionsetup[subfigure]{labelformat=empty}
    \centering
    \subfloat[Bray map]{
    \label{fig:dlab_bray_railroads_detailed_results}
    \includegraphics[angle=270,width=0.25\linewidth]{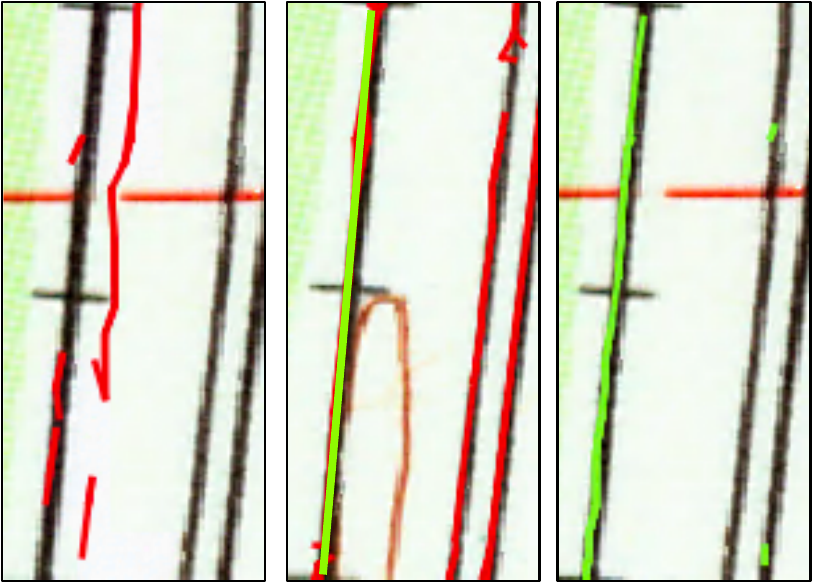}}
    \subfloat[Bray map]{
    \label{fig:dlab_bray_waterlines_extraction}
    \includegraphics[width=0.3\linewidth]{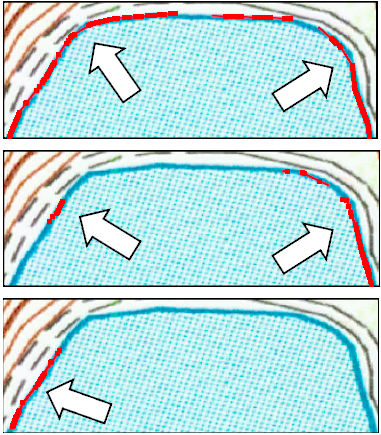}}
    % \hspace{0.15cm}
    \subfloat[Louisville map]{
    \label{fig:dlab_louisville_railroads_extraction}
    \includegraphics[width=0.38\linewidth]{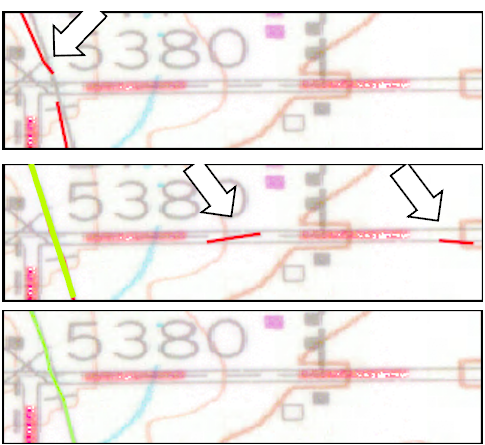}}
    \caption{The recognition visualization. The figures from top to bottom represent the results using the annotations from the original vector, vector-to-raster, and LCA group, respectively. The green and red lines represent the true positive and false positive recognition results, respectively.}
\end{figure}

\begin{figure}[h]
\centering
	\begin{subfigure}{.15\textwidth}
		\frame{\includegraphics[angle=270,width=\linewidth]{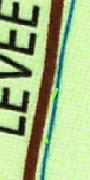}}
		\caption{\small Vec2raster}
		\label{fig:dlab_tp_bray_waterlines_orig_ijgis}
	\end{subfigure}%
	\vspace{0.00mm}
	\begin{subfigure}{.15\textwidth}
		\frame{\includegraphics[angle=270,width=\linewidth]{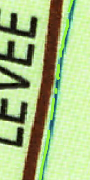}}
		\caption{\small LCA}
		\label{fig:dlab_tp_bray_waterlines_align_ijgis}
	\end{subfigure}%
	\vspace{0.00mm}
	\begin{subfigure}{.15\textwidth}
		\frame{\includegraphics[angle=270,width=\linewidth]{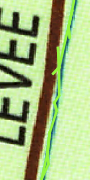}}
		\caption{\small With Ncut loss}
		\label{fig:ncut_tp_bray_waterlines_align_ijgis}
	\end{subfigure}
\caption{The waterlines recognition results in the Bray map. The green lines represent the true positive recognized waterlines. For the annotations, (b) and (c) used LCA, while (a) used the vector-to-raster. For the recognition models, (a) and (b) are results from the model without the normalized cut loss, while (c) used the normalized cut loss. }
\label{fig:completeness_waterlines}
\end{figure}

\subsubsection{With \& Without the Normalized Cut Loss}
\begin{figure}[h]
\centering
	\begin{subfigure}{.22\textwidth}
		\frame{\includegraphics[width=\textwidth]{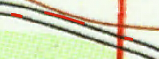}}
		\caption{\small Without normalized cut loss}
		\label{fig:no_ncut_tp_railroads}
	\end{subfigure}
%     \hfill
	\begin{subfigure}{.22\textwidth}
		\frame{\includegraphics[width=\textwidth]{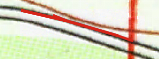}}
		\caption{\small With normalized cut loss}
		\label{fig:ncut_tp_railroads}
	\end{subfigure}
\caption{The false recognized railroads (red lines) in the Bray map. (a) and (b) are the results from the model without and with the normalized cut loss, respectively.}
\label{fig:ncut_fp_exmaple}
\end{figure}

APLS for the railroads and waterlines in the Bray map in Table ~\ref{tab:recognition} show that the Ncut loss helps improve the continuity of recognition results. The APLS with the Ncut loss using LCA is $12.40\%$ higher than without the Ncut loss using LCA. The extracted waterlines (the green lines) using the Ncut loss in Figure~\ref{fig:ncut_tp_bray_waterlines_align_ijgis} did not have gaps, while Figure~\ref{fig:dlab_tp_bray_waterlines_align_ijgis} shows that the results without the Ncut loss were broken. The Ncut loss encourages the nearby pixels with similar colors to be in the same class. Therefore, the Ncut loss reduced the gaps in the extraction results. However, the Ncut loss also caused more false positive extractions close to the target objects and have similar colors as the target objects. The Ncut loss averagely decreased $29.93\%$ correctness. Figure~\ref{fig:ncut_fp_exmaple} shows that the model with the Ncut loss had more continuous false railroads extraction (red lines) in the Bray map. Therefore, the Ncut loss increases the continuity of both true positive and false positive results. 

The low correctness and completeness for the original vector annotations in Table ~\ref{tab:recognition} shows that the Ncut loss could not deal with our annotation noise problem, directly. The annotations from the original vector data had an averagely $60.94\%$ $F_{1}$ score, which shows that the annotation noises are not minor. The major annotation noises are background areas surrounding the target objects. The annotation noises also have high similarities to each other. The Ncut loss cannot correct the annotation noises by comparing the similarity among annotated pixels. Therefore, the Ncut loss cannot directly apply to our recognition task with the noisy labels.

\vspace{-5pt}
\section{Related Work}
\vspace{-1pt}
Existing work~\cite{building_two_stage} uses OSM data to generate the annotations for buildings in satellite imagery. To correct the misaligned annotations from the OSM data, the method leverages a small set of manually corrected annotations to train an alignment correction network (ACN) before training a semantic segmentation model. ACN learns the translation relationships between the buildings' annotations from the external vector data and the buildings in the overhead satellite imagery. During the inference phase, ACN snaps the annotations to the buildings in the maps. ACN assumes that one building's annotation from the vector data corresponds to a building in the maps. Therefore, ACN can only deal with misaligned annotations but cannot handle false annotations. 
    
The existing work~\cite{alignment,road_alignment1, road_alignment2, road_alignment3} about alignment between raster images and vector data may help correct misaligned and false annotations. The existing methods~\cite{road_alignment1, road_alignment2, road_alignment3} focus on aligning the road vector data and roads in the maps. The road alignment algorithms propose diverse methods to detect and match the road intersections for the alignment. However, the waterlines and railroads vector data does not have as many intersections as roads. Therefore, the existing methods for road alignment cannot apply to our label correction problem. The vector-to-raster alignment algorithm~\cite{alignment}, the baseline in the experiment, has a reward function to guide the vector data moving to the target objects in the maps. The reward function places the vector data at the pixels in the target objects' color range in the maps. However, the vector data may be placed at the target objects' edges since edge colors are also in the color range. Consequently, annotations from the vector data may falsely annotate background areas surrounding the target objects.

\vspace{-5pt}
\section{Discussion and Future Work}
\vspace{-0.1cm}
The paper presented a fully automatic geospatial object recognition system. Our system uses the external vector data to automatically generate the training data and thus removes the manual annotating efforts. LCA in the system reduces the annotation noises from the external vector data and thus helps improving the recognition results. The experiments show that our system achieved 0.83 correctness and 0.84 completeness on average, but the average APLS was 0.51. Improving APLS is a priority for future work. The low APLS indicates that the recognized lines have gaps. To reduce the gaps in the recognition results, we will enable the recognition model to explicitly learn that the linear geospatial objects are long continuous lines. The model's recognition will be desired linear objects without gaps. 

\vspace{-5pt}
\section{Acknowledge}
\vspace{-0.1cm}
This material is based upon work supported in part by the National Science Foundation under Grant Nos. IIS 1564164 (to the University of
Southern California) and IIS 1563933 (to the University of Colorado at Boulder), NVIDIA Corporation,  the National Endowment for the Humanities under Award No. HC-278125-21, and the University of Minnesota, Computer Science \& Engineering Faculty startup funds.

\balance
\bibliographystyle{IEEEtran}
\bibliography{egbib}

\end{document}